\documentstyle[aaspp4]{article}

\begin{document}

\title{NEUTRAL POINTS OF OSCILLATION MODES ALONG EQUILIBRIUM SEQUENCES OF RAPIDLY ROTATING 
POLYTROPES IN GENERAL RELATIVITY \\ --- APPLICATION OF THE COWLING 
APPROXIMATION  --- }

\author{
{\large S}{\footnotesize HIN'ICHIROU} {\large Y}{\footnotesize OSHIDA}
\altaffilmark{1} 
	{\footnotesize AND}
{\large Y}{\footnotesize OSHIHARU} {\large E}{\footnotesize RIGUCHI}}
\affil{Department of Earth Science and Astronomy,
	Graduate School of Arts and Sciences,
	University of Tokyo,
	Komaba, Meguro-ku, Tokyo 153, Japan}
\authoremail{yoshida@valis.c.u-tokyo.ac.jp}
\altaffiltext{1}{Research Fellow of the Japan Society for the Promotion of 
Science}

\begin{abstract}

Relativistic Cowling approximation in which all metric perturbations are 
omitted is applied to non-axisymmetric infinitesimal oscillations of 
uniformly rotating general relativistic polytropes.

Frequencies of lower order f-modes which are important in analysis of secular 
instability driven by gravitational radiation are investigated and neutral 
points of the mode along equilibrium sequences of rotating polytropes are 
determined. Since this approximation becomes more accurate as stars are 
more relativistic and/or as they rotates more rapidly, we will be able to 
analyze how a rotation period of a neutron star may be limited by this 
instability.

Possible errors in determining neutral points caused by omitting metric 
perturbations are also estimated.

\end{abstract}


\section{Introduction}

Since the discovery of non-axisymmetric secular instability of 
rotating stars driven by gravitational radiation (\cite{CH,FS,FR1}), 
its possibility to limit a rotational period of a neutron star has been 
pointed out by several authors. As the instability sets in at 
{\it neutral points} in the absence of viscosity, 
it would become the main issue to determine
these neutral points for rapidly rotating stars in general relativity. 
However, thus far, only several investigations in Newtonian gravity 
(\cite{MG,IFD}) and one post-Newtonian investigation (\cite{CL}) have 
been made, because it has been very difficult to manage general 
relativistic perturbations of rotating stars completely.

In such a situation, Yoshida \& Kojima~(1997) have examined the accuracy of 
the relativistic version of the Cowling approximation for oscillations 
of slowly rotating relativistic stars. Comparing the results by the {\it 
relativistic Cowling approximation} 
(hereafter RCA; for earlier literatures, see e.g.
\cite{MVS,FN,LS,IL}) with `exact' 
eigenfrequencies obtained by fully relativistic analysis, they argued that 
the approximation works well for slowly rotating relativistic stars. 
This suggests that we can determine with a fairly good accuracy the neutral 
points by this approximation.

We here investigate the ``counter-rotating'' f-mode oscillations of rapidly 
rotating relativistic stellar models by the RCA. Though, as seen in the 
Newtonian results (\cite{RH}), errors of f-mode eigenfrequencies are larger 
than those of p- or g- modes, absolute values of errors arising from the RCA 
are likely to become smaller for {\it relativistic} and {\it rotating} stars
compared to the Newtonian values (see below). Furthermore we will estimate
roughly the amount of the corrections to the results of the RCA. 

\section{Basic Equations and Numerical Method}

Here we investigate only {\it uniformly rotating polytropic} 
stars. Polytropic relation is defined by $(p/\epsilon_c) = \kappa 
(\epsilon/\epsilon_c)^{1+1/N}$, where $\epsilon_c$ is the central 
energy density, $\kappa$ is a measure of strength of gravity of the star 
and other symbols have their usual meanings. 

The stationary axisymmetric equilibrium states are obtained by the KEH 
scheme (Komatsu, Eriguchi, \& Hachisu 1989). The space coordinates 
$(r, \theta , \varphi)$ and the time coordinate $t$ are used in this 
paper. 
As a parameter characterizing the stellar rotation we adopt the ratio of the
rotational energy to the gravitational energy $T/|W|$ (see \cite{KEH}). 

In the present study, all the Eulerian perturbations of the metric 
components are omitted. Thus our basic equations consist of (1) the 
perturbation equation of the baryon number conservation and (2) the 
perturbation equation of the energy momentum conservation. The perturbed 
Eulerian variables are assumed to have harmonic dependences on the time and 
the azimuthal angular coordinate: $\sim e^{-i\sigma t+im\varphi}$, where 
$m$ is an integer and $\sigma$ is the frequency. The boundary condition of 
the problem is that {\it the Lagrangian perturbation of the pressure 
vanishes at the original stellar surface}. Since the basic equations 
closely resemble those of Newtonian gravity, they are solved by the 
same scheme developed for the Newtonian perturbations 
(\cite{YE}).
\footnote{Although Ipser \& Lindblom~(1992) have presented an elegant 
scalar potential formalism, we do not use that formalism here 
because we prefer a formulation which is parallel to that for the 
Newtonian case (Yoshida and Eriguchi 1995). We thus derive the 
relativistic version of the linearized continuity equation and of 
three components of the linearized Euler equation and solve them directly. 
See appendix 
for explicit form of the equations. }

\section{Results}

\subsection{Test Calculations}

We have checked our RCA code by applying it to two cases, i.e. spherical
stars in general relativity and rotating stars in Newtonian gravity. 
First our frequencies for relativistic spherical stars agree well with those
of other RCA calculations of Yoshida \& Kojima~(1997). As for rapidly rotating
Newtonian stars, by comparing our results with those of the Newtonian Cowling 
approximation which are obtained by the modified code of Yoshida \& 
Eriguchi~(1995), we have found that our RCA code can reproduce the neutral 
points of f-modes with enough accuracy, i.e. within errors of 2\%.

\subsection{Neutral Points along Sequences}

From intersection of $T/|W| - \sigma$ curves of equilibrium sequences and 
$\sigma=0$ we can obtain neutral points which are tabulated in Table 1.

In the framework of Newtonian gravity critical values of $T/|W|$ for the 
$m=2$ `bar' mode have been considered to be universal, i.e. $T/|W|\sim 0.14$ 
irrespective of compressibility and/or the rotation laws (\cite{TJ}). 
However, even for $N=0.5$ polytropes we could not find a neutral point against
this bar mode in weak gravity limit, i.e. for small $\kappa$. This may results
from the crudeness of the Cowling approximation for lower order modes. In the 
Newtonian Cowling approximation, smaller frequencies are obtained than those
calculated from the full treatment. As a consequence, the critical values of
$T/|W|$ at neutral points tend to be raised in this approximation. Thus 
even if $N<0.808$ neutral points for the bar mode may not be reached before 
the termination of equilibrium sequence due to mass-shedding.

On the other hand, neutral points for the bar mode could be found for stronger 
gravity, although the value of $T/|W|$ does not coincide with the Newtonian 
universal value $0.14$. Remarkable is that even for soft equations of state
($N>0.808$) with sufficiently strong gravity, neutral points for the bar mode
do appear. This is not, however, so surprising because general relativistic 
effect on equilibrium states becomes significant.
\footnote{After submitting this paper, we received a preprint 
of Stergioulas \& Friedman (1997), where they also report to 
have obtained 
neutral points of $m=2$ mode 
for softer equations of state 
with taking metric perturbations into consideration. }

Our numerical computational results can be summarized as follows: 
(a) for sequences with $N$ and $\kappa$ kept fixed, values of 
$T/|W|$ at neutral points become smaller as the value of $m$ increases;
(b) for sequences with $N$ and $m$ kept fixed, values of 
$T/|W|$ at neutral points become smaller as the value of $\kappa$ increases;
(c) for sequences with $m$ and $\kappa$ kept fixed, values of 
$T/|W|$ at neutral points becomes smaller as the value of $N$ increases.

In Newtonian gravity, normalized eigenfrequency scales as: 
$\frac{\sigma}{\sqrt{4\pi\rho_c}} \sim \sqrt{l} \frac{\bar{\rho}}{\rho_c}$,
where $\rho_c$ is the mass density at the centre, $\bar{\rho}$ is the mean 
density of the star and $l$ is the zenithal quantum number. From this relation,
normalized frequencies become smaller for models with higher mass 
concentration toward the centre. Because of strong gravity, matter 
distributions are concentrated towards the central region for relativistic 
models. In other words, relativistic stars can be considered to become softer
effectively. Consequently frequencies become lower for larger values 
of $\kappa$.

\section{Discussion -- How Far is the RCA Reliable ?}

Before applying the RCA to realistic neutron stars, it is important to note 
how far this approximation can be reliably applied. In particular we need 
to estimate how the strength of relativity and the amount of rotation affect 
the results. 

It should be noted that eigenfunctions of energy density perturbation for 
f-modes are peaked near the surface.
It implies that surface 
regions are important for f-modes. On the other hand, as is known in Newtonian 
models, relative importance of gravitational perturbations reduces as the 
amount of mass participating in oscillations decreases. 

Consequently relativistic gravity can be considered to improve 
the approximation because the effective concentration of the matter due 
to relativity makes the role of metric perturbations less important. 
Furthermore perturbed density distribution of rotating models is 
more sharply peaked toward the surface of the star as the stars 
are spun up. Therefore the accuracy of the Cowling approximation 
can be relatively increased for rotating models than spherical ones. 
Thus these two factors can be said qualitatively to have advantages 
in deciding neutral points.

However, since there have been no investigations on perturbations of rotating 
stars in full general relativity, the influence of rotation on the errors due 
to the RCA cannot be quantitatively estimated by direct comparison with the 
results of full calculations.  Thus we need some device to have quantitative 
estimation about the errors.

Our proposal is as follows. We assume that qualitative dependence of the 
mode on stellar rotation in general relativity is similar to that of
the Newtonian counterpart. In Newtonian theory, differences between the 
eigenfrequencies obtained by the Cowling approximation and those by the full 
analysis including gravitational perturbations decrease monotonically as stars 
are spun up. If this tendency is the same for general relativistic rotating 
stars, the maximal error of the RCA in determining neutral points can be 
estimated.

Let us select one $T/|W| - \sigma$ curve for a certain equilibrium sequence
computed by the RCA. Since the maximum error of this curve is assumed to
be that of the spherical model, the amount of the maximum error can be 
calculated by comparing our results with exact values of spherical
models in general relativity. A parallel displacement of this  $T/|W| - 
\sigma$ curve by this amount of the maximum error yields a curve with the 
maximal error in determination of neutral points (Fig.~1). Since the amount of 
displacement required to obtain the correct eigenfrequency from the value 
by the Cowling approximation decreases as the value of $T/|W|$ is increased, 
the `true eigenfrequency curve' with gravitational perturbations passes 
between the curves of the Cowling approximation and of the curve of the 
maximal error defined above. 

Therefore if the error of the eigenfrequency for the spherical model and 
the inclination of $T/|W|-\sigma$ curve of the RCA at the approximated 
neutral point are known, the maximal error in determining a neutral point 
will be determined by: 
\begin{equation}
\mbox{Maximal Error of } \frac{T}{|W|} 
= \Delta\sigma_0 \cdot \frac{d\left(\frac{T}{|W|}\right)}{d\sigma} \, ,
\end{equation}
where $\Delta \sigma_0$ is the maximum error of the frequency.

For $N=1$ polytropes we have obtained the maximal errors of $T/|W|$
at neutral points for several modes (Fig.~2). We also display the 
post-Newtonian values of $T/|W|$ at neutral points computed by Cutler 
\& Lindblom (1992). Two results agree within the maximal errors estimated 
here except for $m=2$ modes which are not available for the post-Newtonian 
calculations. 

\section{Concluding Remarks}

Here we have shown that the RCA works rather well in evaluating  
eigenfrequencies of rotating stars in general relativity despite its 
naiveness. As eigenmode analysis in the RCA is much easier than in 
the post-Newtonian treatment, this approximation can be extensively used 
in determining oscillatory frequencies of realistic neutron stars and/or 
determining their neutral points. 

The main drawback of this approximation is that it is hard to include the 
effect of gravitational radiation which is responsible for damping or 
growing of corresponding modes. The post-Newtonian treatment can, in principle,
systematically include this effect by increasing further the order of 
approximation. In contrast the RCA in itself cannot include its effect 
consistently. Some modification such as an estimation of effective damping 
due to gravitational radiation by using eigenfunctions obtained by the RCA 
may be needed.

\acknowledgments

We are grateful to Dr. Shijun Yoshida for discussions. 
We would also like to thank the anonymous referee for letting us know 
the investigation by Stergioulas and Friedman, and Prof. John Friedman for 
sending us a copy of their preprint. 

This paper is presented as a part of the Ph.D. thesis of the first 
author to the Department of Astronomy, University of Tokyo. 
This research has been supported by Grant-in-Aid for Scientific 
Research of the Japanese Ministry of Education, Science and Culture. 


\appendix
\section{Basic Equations of the RCA}

\subsection{Adiabatic perturbations to stationary axisymmetric configurations}
Background spacetime is expressed by the following metric which is used
for stationary axisymmetric spacetime: 
\begin{equation}
 ds^2 = -e^{2\nu(r,\theta)}dt^2 + e^{2\alpha(r,\theta)}(dr^2+r^2d\theta^2)
  + e^{2\beta(r,\theta)}r^2\sin^2\theta(d\varphi-\omega(r,\theta) dt)^2.
\end{equation}
As we adopt the Cowling approximation here, perturbations to this metric 
are totally omitted. 

As for the matter variables, perfect fluid is assumed:
\begin{equation}
	T_{ab} = (p+\epsilon)u_au_b + pg_{ab},
\end{equation}
where $u^a, \epsilon$, and $p$ are the 4-velocity, the energy density and 
the pressure of the fluid, respectively. The matter in the equilibrium
state is assumed to be polytrope, i.e. $p = K\epsilon^{1+1/N}$ where
$K$ is a constant.  Perturbations for the matter are assumed to be adiabatic 
so that the following Lagrangian relation holds for the energy density 
perturbation $\Delta \epsilon$ and the pressure perturbation $\Delta p$:
 \begin{equation}
  \frac{\Delta p}{p} 
   = \gamma \frac{\Delta\epsilon}{\epsilon},
  \end{equation}
where $\gamma$ denotes the adiabatic exponent. Though this exponent 
needs not, in general, coincide with the equilibrium polytropic exponent, 
we assume in this paper that they coincide with each other for simplicity. 
Under this assumption, this Lagrangian relation is reduced to the 
Eulerian one for the Euler perturbations for the energy density 
$\delta \epsilon$ and the pressure $\delta p$ as follows:
 \begin{equation}
  \frac{\delta p}{p} 
   = \gamma\frac{\delta\epsilon}{\epsilon}.
  \end{equation}

Since the background spacetime has timelike and spacelike killing vectors 
$\frac{\partial}{\partial t}$ and $\frac{\partial}{\partial\varphi}$, 
all coefficients which appear in the perturbed equations do not
depend on neither $t$ nor $\varphi$. Consequently solutions of the 
perturbed equations have dependence of $\sim\exp(-i\sigma t +im\varphi)$
on $t$ and $\varphi$ variables where $m$ is an azimuthal eigenvalue and 
integer. 

\newcommand{\ul}[1]{\underline{#1}}
\subsection{Introduction of dimensionless variables}
We introduce the following dimensionless variables by using $\epsilon_c$ 
and $\kappa$ (variables underlined are dimensionless):
\begin{eqnarray}
& & \ul{\epsilon} \equiv \epsilon/\epsilon_c \ \ , \\
& & \ul{p} \equiv p/\epsilon_c
=K\epsilon_c^{1/N}\cdot\ul{\epsilon}^{1+1/N}
\equiv \kappa\ul{\epsilon}^{1+1/N} \ \ , \\
& & \ul{v}^i\equiv u^i/(u^0 \sqrt{\kappa}) \ \ , \\
& & \ul{\sigma}\equiv \sigma/\Omega_c \ \ , \\
& & \ul{\Omega}\equiv \Omega/\Omega_c \ \ , \\
& & \ul{r} = r/r_c \ \ ,
\end{eqnarray}
where $\Omega$ is the angular velocity. Parameters $r_c$ and $\Omega_c$ 
are defined by:
\begin{eqnarray}
& & \Omega_c \equiv \sqrt{4\pi\epsilon_c} \ \ , \\
& & r_c \equiv \sqrt{\kappa}/\Omega_c \ \ .
\end{eqnarray}
The quantity $\sqrt{\kappa}$ is a dimensionless sound velocity at the 
center (normalized by speed of light which is now chosen as $c=1$). 
This parameter measures the system's 
strength of gravity and corresponds to the expansion 
parameter in weak gravity limit.

Further we introduce the {\it Emden function} $\psi$, for it will 
help us to impose the numerical boundary condition at 
the stellar surface properly: 

\begin{equation}
 \ul{\epsilon} \equiv \psi^N \quad ,\quad \ul{p} = \kappa\psi^{N+1} \ \ .
\end{equation}

\subsection{Surface-fitted coordinates}
Since a rapidly rotating star deforms from spherical configuration, 
grid points do not, in general, fall on the surface in the ordinary 
polar coordinates. This makes it difficult to impose the boundary 
condition at the stellar surface. To avoid this difficulty, we 
introduce the {\it surface-fitted coordinates} as in the Newtonian 
analysis (see e.g. \cite{YE}). For a configuration with the stellar 
surface $r = R_s(\theta)$ in the polar coordinates $(r, \theta)$, 
new coordinates $(\tilde{r}, \tilde{\theta})$ are defined by
\begin{equation}
 \tilde{r} \equiv r/R_s(\theta) \ \ ,
\qquad\qquad \tilde{\theta} = \theta \ \ .
\end{equation}
Although owing to this coordinate transformation the basic equations 
become slightly complicated, the boundary condition at the surface,
i.e. vanishing of the Lagrangian perturbation of the pressure 
\begin{equation}
	\Delta p = 0,
\end{equation}
can be imposed easily.

Hereafter, for simplicity, 'tilde' and underlines will be omitted in 
the equations expressed by this new coordinate.

\newcommand{\dr}[1]{\partial_r#1}
\newcommand{\dt}[1]{\partial_{\theta}#1}
\newcommand{\dts}{(\partial_{\theta}-rS\partial_r)}
\subsection{Hydrodynamic Equations in the Cowling approximation}
Basic equations of relativistic hydrodynamics are
\begin{eqnarray}
& & \nabla\cdot(n u) = 0 \ \ , \quad \mbox{(baryon number conservation)}\\
& & \nabla_bT^{ab} = 0 \ \ , \quad \mbox{(energy-momentum
 conservation)}\\
& & \tilde{d}\epsilon = \frac{\epsilon +p}{n}\tilde{d}n \ \ ,
\quad \mbox{(the first law of thermodynamics)}
\end{eqnarray}
where $n$ is the baryon number density, and differential operators 
$\nabla$ and $\tilde{d}$ are the covariant derivative and the exterior 
derivative, respectively.

By introducing 3-velocity components on orthogonal basis,
\begin{eqnarray}
 U &\equiv& -i e^{\alpha-\nu} \delta v^r,\\
 V &\equiv& -i re^{\alpha-\nu} \delta v^{\theta},\\
 W &\equiv& e^{\beta-\nu}r\sin\theta \delta v^{\varphi},
\end{eqnarray}
and rewriting the variable $\delta \psi$ as $\Psi = \delta\psi$, 
the basic equations for the perturbed quantities are written as follows.

For the baryon number conservation, we have,
\begin{eqnarray}
(\sigma-m\Omega)\Psi &+& \left[-\frac{\psi
(1+\kappa\psi)}{NR_s(\theta)}e^{\nu-\alpha}\right]\dr{U}
+ \left[-\frac{e^{\nu-\alpha}\dr{\psi}}{R_s(\theta)} + \frac{\psi
(1+\kappa\psi)e^{\nu-\alpha}\dr{\psi}}{2NAR_s(\theta)}\right. \nonumber\\
&&\left. -\psi(1+\kappa\psi)\frac{e^{\nu-\alpha}}{NR_s(\theta)}
\left(2\dr{\nu} + \dr{\alpha} + \dr{\beta} + \frac{2}{r}\right)
\right]U\nonumber\\
&+& \left[-\frac{\psi(1+\kappa\psi)e^{\nu-\alpha}}
{NR_s(\theta)r}\right]\dt{V} + \left[-\frac{\psi
(1+\kappa\psi)e^{\nu-\alpha}S}{NR_s(\theta)r}\right]
\dr{V}\nonumber\\
&+& \left[-\frac{e^{\nu-\alpha}(\dt{\psi}-rS\dr{\psi})}
{R_s(\theta)r} + \frac{\psi(1+\kappa\psi)e^{\nu-\alpha}
(\dt{A}-rS\dr{A})}{2NAR_s(\theta)r} \right.\nonumber\\
 & & \left. -\frac{e^{\nu-\alpha}\psi
(1+\kappa\psi)}{NR_s(\theta)r}\left((\partial_{\theta}-rS\partial_r)
(2\nu+\alpha+\beta) + \cot{\theta}\right)\right]V\nonumber\\
&+& \left[\frac{\psi(1+\kappa\psi)(\omega-\Omega)
e^{\nu+\beta}R_s(\theta)r\sin\theta}{NA}(\sigma-m\Omega)
\right.
\left. -\frac{\psi(1+\kappa\psi)e^{\nu-\beta}m}
{NR_s(\theta)r\sin\theta}\right]W = 0.
\end{eqnarray}

Relativistic version of the Euler equation is obtained by applying 
the projection tensor $P_{ab} = u_au_b + g_{ab}$ onto the equation of 
energy-momentum conservation as follows:
\begin{eqnarray}
 (\sigma-m\Omega)U &+& \left[r\dr{\omega} 
+ 2(\omega-\Omega)r\dr{\beta} 
+ 2(\omega-\Omega) 
+ 2\kappa(N+1)Br\dr{\psi}(\omega-\Omega)\right]
e^{\beta-\alpha}\sin\theta W\nonumber\\
&-& (N+1)ABe^{-\alpha-\beta}\frac{\dr{\Psi}}{R_s(\theta)}
+ \kappa(N+1)AB^2e^{-\alpha-\nu}\frac{\dr{\psi}}{R_s(\theta)}
\Psi = 0,\\
(\sigma-m\Omega)V &+& \left[\dts\omega 
+ 2(\omega-\Omega)\dts\beta 
+ 2(\omega-\Omega)\cot\theta \right.\nonumber\\
&&\left.+ 2\kappa(N+1)B\dts\psi(\omega-\Omega)\right]
e^{\beta-\alpha}\sin\theta W\nonumber\\
&-& (N+1)AB\frac{e^{-\nu-\alpha}}{R_s(\theta)r}\dts\Psi
+ \kappa(N+1)AB^2\frac{e^{-\nu-\alpha}}{R_s(\theta)r}\dts\Psi = 0,\\
(\sigma-m\Omega)W &-& \left[2(\omega-\Omega)
\dr{\nu}-\kappa(\omega-\Omega)^2e^{2\beta-2\nu}
R_s(\theta)^2r^2\sin^2\theta - \dr{\omega}
-2(\omega-\Omega)\left(\dr{\beta}+\frac{1}{r}\right)\right]
e^{\beta-\alpha}r\sin\theta U\nonumber\\
&-& \left[2(\omega-\Omega)\dts\nu
-\kappa(\omega-\Omega)^2e^{2\beta-2\nu}
R_s(\theta)^2r^2\sin^2\theta\dts\omega \right.\nonumber\\
&&\left. - \dts\omega
-2(\omega-\Omega)(\dt{\beta}-rS\dr{\beta}+\cot\theta)
\right]e^{\beta-\alpha}\sin\theta V \nonumber\\
&+& \kappa\sigma(N+1)AB(\sigma-\Omega)e^{\beta-3\nu}
R_s(\theta)r\sin\theta \Psi\nonumber\\
&+& m(N+1)AB\left[\frac{e^{-\beta-\nu}}{R_s(\theta)r\sin\theta}
-\kappa\omega(\omega-\Omega)e^{\beta-3\nu}
R_s(\theta)r\sin\theta \right]\Psi = 0 \ \ ,
\end{eqnarray}
where functions $A$, $B$ and $S$ are defined as,
\begin{equation}
 A \equiv g_{00} + 2g_{0j}v^j + g_{jk}v^j v^k=-e^{2\nu}
+e^{2\beta}r^2\sin^2\theta(\omega-\Omega)^2,
\end{equation}
\begin{equation}
 B \equiv \frac{1}{1+\kappa\psi},
\end{equation}
\begin{equation}
	S(\theta) \equiv \frac{1}{R_s(\theta)}\frac{dR_s(\theta)}{d\theta}.
\end{equation}
\newpage

\figcaption[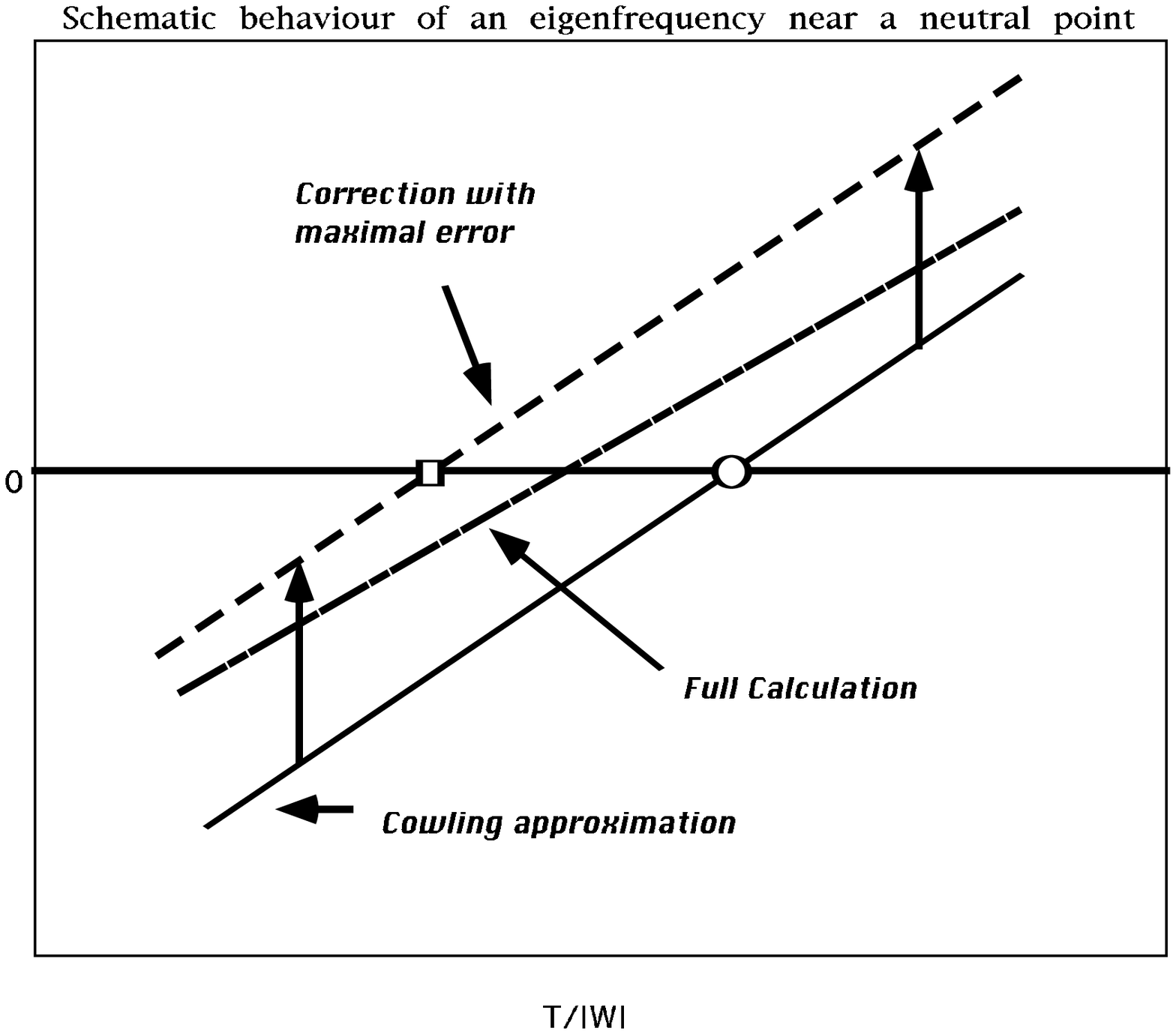]{A schematic behaviour of an eigenfrequency 
near the neutral point. `Full calculation' means values obtained
from full inclusion of gravitational perturbations. `Correction with 
maximal error' line is the line with the maximal error denoted by 
vertical arrows and extrapolated from a spherical model. The open circle 
and square denote approximated neutral points.}

\figcaption[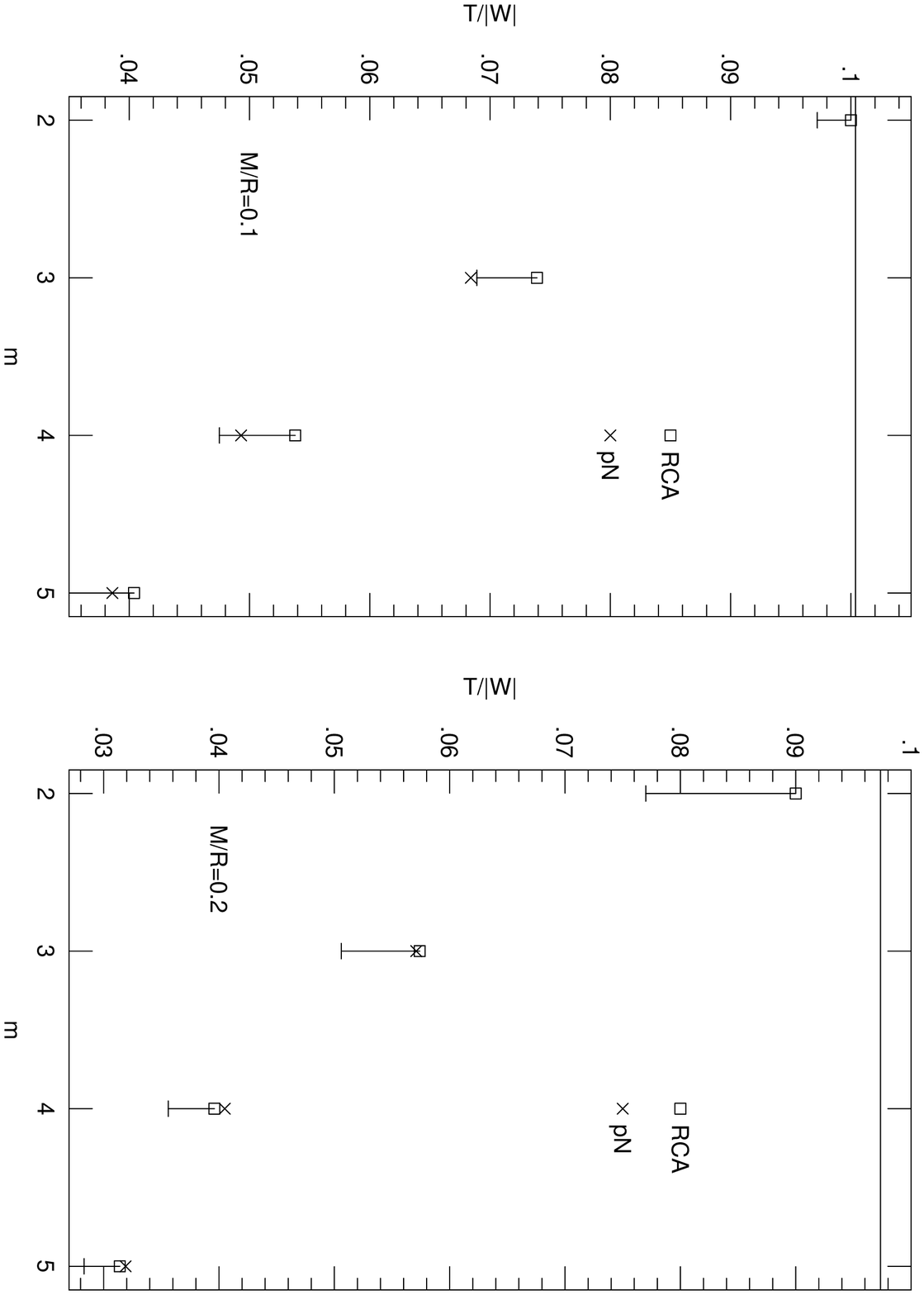]{Critical values of $T/|W|$ at neutral points 
for several orders of f-modes for $N=1$ polytrope. In this figure, 
as a measure of strength of gravity, the ratio of the mass to the radius
in Schwarzschild coordinate $M/R$ evaluated for spherical models is adopted. 
Error bars with open squares for both sides show maximal errors expected 
for the RCA. Integers $m$ are the orders of the mode. Crosses are values of 
post-Newtonian results.
Horizontal lines in the upper region denote critical values where equilibrium 
sequences terminate due to mass-shedding. }

\begin{deluxetable}{cccccccccccc}
\scriptsize
\tablecolumns{12}
\tablewidth{0pc}
\tablecaption{$T/|W|$ Values at Neutral Points}
\tablehead{
\colhead{}    &  \multicolumn{3}{c}{$N=0.5$} &   \colhead{}   & 
\multicolumn{3}{c}{$N=1.0$} &   \colhead{}   & \multicolumn{3}{c}{$N=1.5$}\\
\cline{2-4} \cline{6-8} \cline{10-12}\\
\colhead{$\kappa$} & \colhead{$m=2$}   & \colhead{$m=3$}    & \colhead{$m=4$} 
& \colhead{}    & \colhead{$m=2$}   & \colhead{$m=3$}    & \colhead{$m=4$} & 
\colhead{}    & \colhead{$m=2$}   & \colhead{$m=3$}    & \colhead{$m=4$}  }
\startdata
$0.001$ & \nodata & $1.36E-1$ & $9.06E-2$ & & \nodata & $9.31E-2$ & $7.90E-2$ & & \nodata & $5.69E-2$ & $4.92E-2$ \nl
$0.01$ & \nodata & $1.28E-1$ & $8.30E-2$ & & \nodata & $9.02E-2$ & $7.39E-2$ &
 & \nodata & $5.66E-2$ & $4.84E-2$ \nl
$0.1$ & $1.47E-1$ & $9.25E-2$ & $6.17E-2$ & & $9.70E-2$ & $6.95E-2$ & $4.91E-2$ & & $5.65E-2$ & $4.77E-2$ & $3.44E-2$ \nl
$0.5$ & $7.92E-2$ & $5.11E-2$ & $3.75E-2$ & & $7.53E-2$ & $4.11E-2$ & $2.93E-2$ & & $4.14E-2$ & $2.92E-2$ & $1.98E-2$ \nl
\enddata
\end{deluxetable}

\end{document}